\documentclass[prb,twocolumn,showpacs,floatsfix]{revtex4-1}

\usepackage{graphicx,epic,eepic}% Include figure files
\usepackage{bm}% bold math
\usepackage[]{graphics,epsfig,ams,amssymb,color}
\newcommand{\be}{\begin{equation}}
\newcommand{\ee}{\end{equation}}
\begin{document}
\title{Stochastic heating of a molecular nanomagnet}
%Manuscript Title:\\with Forced Linebreak}% Force line breaks with \\
%\title{Dynamics of dipolar molecular chain in crossed static: Soliton formation}
%Manuscript Title:\\with Forced Linebreak}% Force line breaks with \\

\author{L. Chotorlishvili$^{1,3}$, P. Schwab$^{1}$, Z. Toklikishvili$^{3}, $~J. Berakdar$^{2}$}
%%
%\institute{
%  \inst{1} Institut f\"ur Physik, Martin-Luther Universit\"at Halle-Wittenberg, Heinrich-Damerow-Str.4
%06120 Halle, Germany\\
%\inst{2}Institute f\"ur Physik, Universit\"at Augsburg, 86135 Augsburg, Germany\\
% \inst{3} Physics Department of the Tbilisi State University,
%                 Chavchavadze av.3, 0128, Tbilisi, \\ Georgia }
%  \inst{2} Second Institute - Address

\affiliation{$^1$Institut f\"ur Physik, Universit\"at Augsburg, 86135 Augsburg, Germany\\
   $^2$Institut f\"ur Physik, Martin-Luther Universit\"at
Halle-Wittenberg, Heinrich-Damerow-Str.4, 06120 Halle, Germany \\
  $^3$Physics Department of the Tbilisi State University,
                 Chavchavadze av.3, 0128, Tbilisi, Georgia}
  \begin{abstract}
  We study the excitation dynamics of a single molecular nanomagnet by static and pulsed magnetic fields.
  Based on a stability analysis of the classical magnetization dynamics we identify analytically the
    fields parameters for which the energy is stochastically pumped into the system in which case
     the magnetization
   undergoes  diffusively and irreversibly a large angle deflection. An approximate
   analytical expression for the diffusion constant
   in terms of the fields parameters is given and assessed by full  numerical calculations.
  \end{abstract}
  %Valid PACS numbers may be entered using the \verb+\pacs{#1}+
%command.
\maketitle
\section{Introduction}
In molecular nanomagnets (MNMs) \cite{MM} such as  in Mn$_{12}$ acetates
 the magnetic core of  the molecule is surrounded by organic non-magnetic
 ligands that extinguish  the inter MNM exchange interactions.
 Hence, much of the on-site physical properties are deducible by studying
 a single MNM, albeit the dipolar interaction is present and
is essential for ordering phenomena \cite{garanin}.
% by
%Interesting feature of magnetic materials is the fact that large
%number of individual paramagnetic spins may form macroscopic
%magnetic moment in the bulk leading to the superparamagnetism.
%Phenomenon by which magnetic materials may exhibit a behavior
%similar to paramagnetism at temperatures below the Curie or the
%Neel temperature \cite{Magnetic}. Developing of efficient tools
%for manipulating  of large macroscopic spins in a controlled
%manner is a striking problem of spintronics \cite{Introduction}
Characteristic for MNMs  is the relatively large effective spin
(e.g., $S$ = 16 for  Mn$_{12}$) and the magnetic anisotropy
\cite{MM}. MNMs exhibit a  series of  phenomena \cite{MM} that are
relevant for applications in spintronics and quantum information
 \cite{Mabuchi}; most notably is the bistability behaviour, the
 resonance  tunneling of magnetization \cite{Chudnovsky} and
  the large spin relaxation time.
The dynamical control and switching of the magnetization via external fields is
a key ingredient on the way to utilizing MNM for technological applications.
 In this context,  the role of thermal and environmental
effects  have been considered \cite{Khapikov,Denisov}. For
a single MNM
\cite{Lis,Quantum,Friedman,Hernandez,Lionti, Chudnovsky, Thomas,
Wernsdorfer,Hennion} at very low temperatures the
 main switching  mechanism is the quantum
tunneling of  magnetization \cite{Quantum}. This is
  due to the large anisotropy  barrier
\cite{Gatteschi}.
For an initially excited MNM  and driven by external magnetic fields
 we have shown recently
  \cite{Chotorlishvili}
  % that new mechanism of magnetization
%switching emerges.  As was shown,
that the phase space of the magnetization
has a rich structure containing a separatrix of topologically different
domains. Switching occurs at the separatrix as
a consequence of a transition between these
domains. However, the issue how to inject energy in this
nonlinear system, i.e. how to realize the initially excited state near the separatrix,
 has not been addressed.
%
%Nevertheless some interesting problems are not addressed
%yet. In particular injecting of energy in the nonlinear system and
%realization of exited state, corresponding to the separatrix
%condition is a crucial issue for dynamically induced switching
Starting from the ground state,
this question is not  answered  by resonant fields as the system
is non-linear, i.e. it changes its eigenfrequency  as
the oscillation amplitude varies
\cite{Chotorlishvili}.
Formally the equations of motion for a single molecular magnet resembles the Landau-Lifshitz-(Gilbert) equation
without the Gilbert  damping. In the present case the negligible damping is an inherent
 system property and not a shortcoming of theory. This difference is insofar important as  without dissipation
a precessional switching, e.g. as proposed in \cite{Bauer}, is not achievable. Hence, 
 new switching schemes
are needed for MNM that are different from those known for magnetic materials. 
One scheme proposed recently \cite{Chotorlishvili} relies on a stochastic, diffusion-type switching.
For this to work however, the system has to be excited to a desired state (near the separatrix). The question of how
to achieve that is still open.
 In this paper we show that using  appropriate polychromatic magnetic pulses
we can achieve a stochastic heating of a molecular magnet as appropriate for the stochastic switching.
 We derive approximate analytical expressions for the field parameters that allow for the stochastic heating and
 test for our analytical predictions with full numerical calculations.
\section{Theoretical model}

We study a single molecular magnet (MM), e.g. Fe$_{8}$ or
Mn$_{12}$ acetates and choose the  $z$ axis to be along the
 uniaxial anisotropy direction (easy axis).
 The MM is subjected to a constant magnetic
field with an amplitude $H_{0}$, applied along the $x$-axis (hard axis) as well as
 to a series of magnetic  pulses  ${\cal F}(t)$ that are linearly polarized in the $x$ direction.
 The Hamiltonian we write as \cite{Chotorlishvili}:
\begin{eqnarray}\label{eq:hamiltonian}
&&\hat{H}=\hat{H}_{0}+\hat{H}_{I}, \\
&&\hat{H}_{0}=-D\hat{S}_{z}^{2}+g\mu_{B}H_{0}\hat{S}_{x},
\hat{H}_{I}=g\mu_{B}{\cal F}(t)\hat{S}_{x}.\nonumber
\end{eqnarray}
Here $D$ is the longitudinal anisotropy constant, $S_{x}$,
$S_{y}$,  and $S_{z}$ are the spin operators projections   along the
$x,y$, and $z$ directions, respectively. $g$ is the Land\'e factor,  and $\mu_{B}$ is
the Bohr magneton. Since the spin of the molecular nanomagnet is quite
large a classical approximation is appropriate. Hence, it is
advantageous to introduce the variables $(S_{z},~\varphi)$ via the
transformation $$S_{x}=\sqrt{1-S_{z}^{2}}\cos\varphi,\;
S_{y}=\sqrt{1-S_{z}^{2}}\sin\varphi$$ and rewrite
(\ref{eq:hamiltonian}) in the compact form \cite{Chotorlishvili}
\begin{eqnarray} \label{eq:mag}
H(t)&=&-\frac{\lambda}{2}S_{z}^{2}+\sqrt{1-S_{z}^{2}}
\cos\varphi-\sqrt{1-S_{z}^{2}}\cos\varphi F(t)\nonumber\\
\lambda&=&\frac{2DS}{g\mu_{B}H_{0}},\:
F(t)=\frac{{\cal F}(t)}{g\mu_{B}H_{0}} . \end{eqnarray}

 Suppose that the applied constant field is weak
$\lambda>1$ and at the initial moment of time the system resides  near to
the ground state $S_{z}(t=0)\approx\pm1$. How to pump efficiently energy
into the system  such that we reach the excited states
 near the separatrix, where then a dynamically induced switching is
realizable? To answer this question we make a transition to the
action-angle variables $(I,\varphi)$ in which the Hamiltonian
(\ref{eq:mag}) reads
\begin{eqnarray}\label{eq:hammol}
&&H=H_{0}(I)+V(I,\varphi)F(t),\nonumber \\
&&H_{0}=\omega(I)I, \omega(I)=\bigg[\frac{d I(\Sigma)}{d \Sigma}\bigg]^{-1}, \\
&&I(\Sigma)=\frac{1}{\pi}\int S_{z} (\Sigma,\varphi)d\varphi,~~~~
\Sigma=-H. \nonumber
\end{eqnarray}
The equations of motion (EOM)  are
\begin{eqnarray}
&&\dot{I}=-\frac{\partial H}{\partial \varphi}=-\frac{\partial V(I,\varphi)}{\partial \varphi}F(t),\nonumber \\
&&\dot{\varphi}=\frac{\partial H}{\partial I}=\omega
(I)+\frac{\partial V(I,\varphi)}{\partial I}F(t).
\end{eqnarray}
In absence of the time dependent perturbation, $I$ is an integral
of motion
($\varphi(t)$ is fast variable, however).
% In addition frequency of
%nonlinear oscillations depends on the amplitude of oscillations
%$\omega(I)$.
%
%Therefore, situation is typical for KAM theory and nonlinear
%resonance \cite{Zaslavsky}.
%
%
For an applied
monochromatic field it is not possible to keep in resonance with $\omega$,
for  $\omega$ depends on $I$  and hence it changes in time. %Suppose that variable
%field is monochromatic and for some values of the action $I=I_{1}$
%resonance condition is hold $\omega(I_{1})=\Omega$. Here $\Omega$
%is a frequency of variable field $F(t)$. Resonance in itself will
%change values of action and $\omega(I_{2})\neq\Omega$. However, if
A polychromatic field offers a wider range of frequencies that
may match the dynamical frequency of the system. %the situation changes
%is applied then there is possibility to
%reach new resonance $\omega(I_{2})=2\Omega$.  All this is will be
To be more concrete let us assume the applied field
to be of the form
 \be \label{eq:pulse}F=\varepsilon_{0} T
\sum\limits_{n=-\infty}^{\infty}\delta_{\tau}(t-nT)=\varepsilon_{0}
\frac{\tau}{T}
\sum\limits_{n=-1/\tau}^{1/\tau}\cos\bigg(\frac{2\pi}{T}nt\bigg),\ee
where $\tau$ is the pulse duration, $T>\tau$ is interval between
pulses, and $\varepsilon_{0}$ is the pulse strength.

\section{Stochastic heating}
Of particular interest for us is the situation of
overlapping resonances which is realized when
\cite{Zaslavsky} \be\label{eq:KA}
 K'=\varepsilon_{0} T I(\Sigma) \frac{d \omega(I)}{d I(\Sigma)}>1,
 \ee
in which case the  dynamics turns
 chaotic  \cite{Zaslavsky}, i.e.
the system jumps from one resonance to other in a random way. A
key point here is the  irreversibility of the dynamics that 
emerges due to nonlinearity and without any thermal effects nor external
random forces. Hence, we expect a  "stochastic heating"  of a MM
 subjected to the pulses (\ref{eq:pulse}) when the criterion  (\ref{eq:KA}) is fulfilled.

%In what follows, general formalism is given in \cite{Zaslavsky}.
%Suppose that series of rectangular pulses are applied on the
%system: \be \label{eq:pulse}F=\varepsilon_{0} T
%\sum\limits_{n=-\infty}^{\infty}\delta(t-nT)=\varepsilon \tau
%\sum\limits_{n=-1/\tau}^{1/\tau}\cos\bigg(\frac{2\pi}{T}nT\bigg),\ee
%where $\tau$ is the pulse width, $T>\tau$ interval between pulses.
%We see that  contains many harmonics and if
%resonance overlapping
% of the system happens.

Assuming that  $S_{z}(t=0)\approx\mp1$ and
$\lambda>1$  we find
\begin{eqnarray}
&&I(\Sigma)=\int S_{z}(\Sigma,\varphi)d\varphi=\frac{2}{\sqrt{\lambda}}\sqrt{\Sigma-1}E\Big(\frac{2}{\Sigma-1}\Big),\nonumber \\
&&\omega (I)=\bigg[\frac{d I(\Sigma)}{d \Sigma}\bigg]^{-1}=\frac{\sqrt{\lambda}\sqrt{\Sigma-1}}{K\big(\frac{2}{\Sigma-1}\big)},\\
&&\frac{d \omega(I)}{d I}=\frac{d \omega(I)}{d \Sigma}\frac{d \Sigma}{d I}=\omega(I)\frac{d \omega(I)}{d \Sigma}, \nonumber \\
&&\frac{d \omega (I)}{d
\Sigma}=\frac{\sqrt{\lambda}\sqrt{\Sigma-1}E\big(\frac{2}{\Sigma-1}\big)}{2(\Sigma-3)K^{2}\big(\frac{2}{\Sigma-1}\big)}.\nonumber
\end{eqnarray}
Thus we infer  \be\label{eq:K} K'=\varepsilon_{0} T
\frac{\sqrt{\lambda}(\Sigma-1)^{3/2}E^{2}\big(\frac{2}{\Sigma-1}\big)}{(\Sigma-3)K^{3}\big(\frac{2}{\Sigma-1}\big)}>1.
\ee  $E\big(\frac{2}{\Sigma-1}\big)$,
$K\big(\frac{2}{\Sigma-1}\big)$ are the complete elliptic
integrals in the notation of Ref.[\onlinecite{Handbook}]. In the regime of chaotic motion, when Eq.(6) holds,
a dynamical description becomes inappropriate. The adequate language for
the study of the magnetization dynamics in this case is an approach
based for example on the Fokker-Planck equation. A Fokker-Planck equation for
the distribution function of the action $f(I,t)$ can be set up in
a similar way as done in Ref.[\onlinecite{Toklikishvili}]:
\begin{eqnarray} \label{eq:Fokker}
\frac{\partial f(I,t)}{\partial t}=\frac{1}{2}\frac{\partial}{\partial I}D(I)\frac{\partial f(I,t)}{\partial I}, \,
D(I)=\frac{\varepsilon^{2}T}{2}\big(1-\frac{2}{\lambda}I\omega(I)\big).\nonumber
\end{eqnarray}
The relevant quantity is the averaged value of the action $\langle
I(t,\Sigma)\rangle_{f}=\overline{I(t)}.$
%Multiplying
%(\ref{eq:Fokker}) from both sides on the  $I$ and
 Using the relations
\be \int I \frac{\partial f(I,t)}{\partial t}dI=\frac{1}{2}\int I
\frac{\partial}{\partial I}D(I)\frac{\partial f(I,t)}{\partial
I}dI, \ee \be \dot{\bar{I}}(t)=-\frac{1}{2}\int D(I)\frac{\partial
f(I,t)}{\partial I}dI=\frac{1}{2}\int\frac{\partial D(I)}{\partial
I}f(I,t)dI, \ee \be \frac{\partial D(I)}{\partial
I}=-\frac{\varepsilon^{2}_{0}T}{2\lambda}\omega(I)\bigg(1+I\frac{d
\omega(I)}{d \Sigma}\bigg), \ee
 we infer that
 \be
\dot{\bar{I}}=-\frac{\varepsilon^{2}_{0}T}{2\lambda}\omega(\bar{I})\bigg(1+\bar{I}\frac{d
\omega(\bar{I})}{d \Sigma}\bigg).\ee 
We note that $I$ and
$\omega(I)$ are functions of $\Sigma$ and make the approximation that \be
\bar{I}(\Sigma)=I(\bar{\Sigma}),~~~
\bar{\omega}(I)=\bar{\omega}\big(I(\bar{\Sigma})\big),
 \ee 
  meaning that  correlations between the 
  random variables are neglected. Formally, we can  
   systematically improve  on this approximation
   by accounting for higher moments for the correlation functions of the random variables.
Neglecting these correlations  we  find for  $\dot{\bar{I}}$ 
\be
\dot{\bar{I}}=\overline{\frac{d I}{d \Sigma}\frac{d \Sigma}{d
t}}=-\frac{\varepsilon^{2}_{0}T}{2\lambda}\omega(\bar{I})\bigg(1+\bar{I}\frac{d
\omega(\bar{I})}{d \Sigma}\bigg),\ee \be \overline{\frac{d
\Sigma}{d
t}}=-\frac{\varepsilon^{2}_{0}T}{\lambda}\omega^{2}(\bar{I})\bigg(1+\bar{I}\frac{d
\omega(\bar{I})}{d \Sigma}\bigg) \ee we conclude that
\be\label{eq:sigma} \frac{d \overline{\Sigma}}{d
t}=-\varepsilon^{2}_{0}T\frac{\overline{\Sigma}-1}{K^{2}(\frac{2}{\overline{\Sigma}-1})}
\Bigg(1+\frac{\overline{\Sigma}-1}{\overline{\Sigma}-3}\frac{E^{2}
(\frac{2}{\overline{\Sigma}-1})}{K^{2}(\frac{2}{\overline{\Sigma}-1})}\Bigg).\ee
>From the asymptotical solution of $\bar{\Sigma}>1$
\be\label{eq:sigma1}
\bar{\Sigma}(t)=\frac{\lambda}{2}\exp\bigg[-\frac{4}{\pi}\varepsilon^{2}_{0}Tt\bigg]\ee
we uncover a diffusive decay of $\bar{\Sigma}(t)$, meaning that the
energy is increased diffusively due to the relation
$H=-\bar{\Sigma}(t)$), albeit eq. (\ref{eq:K}) must be obeyed.
%$\bar{\Sigma}>1$ and beyond eq.(\ref{eq:sigma1}) is not correct
%strictly speaking.
\begin{figure}[t]
 \centering
  \includegraphics[width=6.5cm]{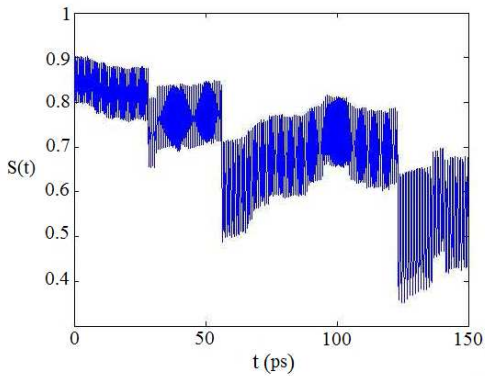}
  \caption{ A demonstration of the stochastic deflection of the magnetization vector from the initial  state $S_{z}(0)=0.9$ upon
  applying a series of rectangular pulses. The results are obtained by solving numerically for the Hamiltonian equations.
   The time scale is set by values of the constant magnetic field $t\rightarrow t/g\mu_{B}H_{0}$ [16]. The parameters are
  $\varepsilon_{0}=0.4$, $T=1ps$, $\tau=0.01ps$, $\lambda=10$. Note, the criterion of the stochasticity, given by  (\ref{eq:K}) is realized since $K'=1.75>1$.} \label{Fig:1}
\end{figure}

\begin{figure}[t]
 \centering
  \includegraphics[width=6.5cm]{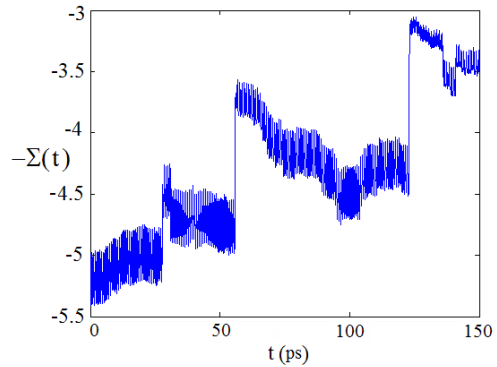}
  \caption{Illustration of the stochastic heating by external driving fields after having started    from the initial state $S_{z}(0)=0.9$. The results are obtained as in Fig. \ref{Fig:1} with the parameters
  $\varepsilon_{0}=0.4$, $T=1ps$, $\tau=0.01ps$, $\lambda=10$.
  The criterion of stochasticity (\ref{eq:K}) is fulfilled ($K'=1.75>1$)
   and a  diffusive increase of the system energy is observed. This numerical result is qualitatively consistent with the analytical prediction, given by Eq. (17). For a better numerical agreement one has to go beyond the approximation (13) and consider higher moments for the correlation functions of the random variables} \label{Fig:2}
\end{figure}

\begin{figure}[t]
 \centering
  \includegraphics[width=6.5cm]{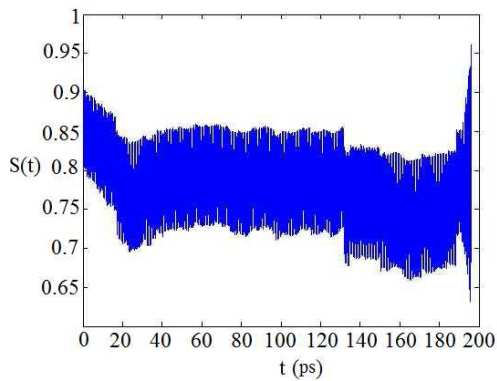}
  \caption{The time evolution of the magnetization vector when starting from the initial  state $S_{z}(0)=0.9$.
  The results are obtained by solving numerically for the Hamiltonian equations with a series of rectangular pulses.  The parameters are
  $\varepsilon_{0}=0.1$, $T=1ps$, $\tau=0.01ps$, $\lambda=10$, $K'=0.43<1$.  The criterion of stochasticity is not fulfilled. The
  orientation of the magnetic vector is not changed significantly,  and  only  fluctuations around the ground state is observed.} \label{Fig:3}
\end{figure}

\section{Numerical results}
The exact numerical results shown in Figs.(\ref{Fig:1}, \ref{Fig:2},
\ref{Fig:3})  evidence that the mechanism of stochastic
heating is indeed present and is quite efficient.

Other scenario for the magnetization control is to employ
 only the
periodic series of rectangular pulses applied  along the hard axis,
i.e. to switch off the static field.
%\be\label{ham}
%\hat{H}(t)=-\frac{DS_{z}^{2}}{2}+H\sqrt{1-S_{z}^{2}}\cos\varphi\cdot
%T\sum\limits_{n=0}^{\infty}\delta(t-n T).\ee Here $H$ is the
%amplitude of pulses and $T$ is the time interval between them. Or
Measuring the energy in units of $D$ we write for the scaled
Hamiltonian
%\begin{eqnarray}
%&&
$$\bar{H}(t)={H}(t)/D
=-\frac{S_{z}^{2}}{2}+V(s_{z},\varphi),$$ where $$
V(s_{z},\varphi)=V_{0}(s_{z},\varphi)\,
T\sum\limits_{k=-\infty}^{+\infty}\delta(t-k
T),$$ and $$ V_{0}(s_{z},\varphi)=\varepsilon\sqrt{1-s_{z}^{2}}\cos\varphi,$$ with $$\varepsilon=\frac{\varepsilon_0}{D}.
$$
%\nonumber\end{eqnarray}
The equations of motion
\begin{eqnarray}\label{eq:motion}
&&\dot{\varphi}=\omega(s_{z})+\varepsilon\frac{\partial
V_{0}(s_{z},\varphi)}{\partial
s_{z}}T\, \sum\limits_{k=-\infty}^{+\infty}\delta(t-k T),\\
&&\dot{s}_{z}=-\varepsilon\frac{\partial V_{0}(s_{z},\varphi)}{\partial
\varphi}T\, \sum\limits_{k=-\infty}^{+\infty}\delta(t-k
T),\, \omega(s_{z})=-s_{z}\nonumber
\end{eqnarray}
 can be integrated exactly in this case
by formulating them as recurrence
relations \cite{Zaslavsky} using the evolution
operator $\hat{T}$ that propagate the system
from the time $t_0$ to $T$, i.e.
\begin{figure}[t]
 \centering
  \includegraphics[width=6.5cm]{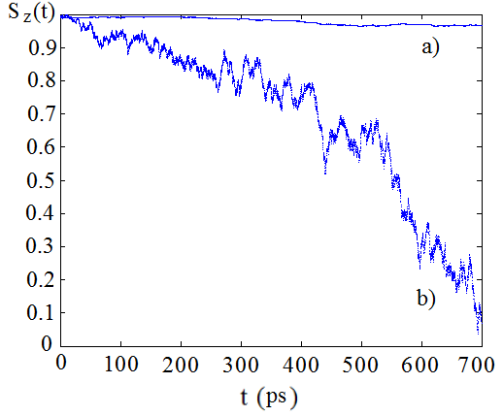}
 \caption{$S_n^z(t)$ as a function of the number of applied kicks $t=nT$ obtained from
 a full numerical integration of the recurrence relations eq.(\ref{green}).
The basic control parameter is the time interval
between the pulses $T$ which allows a tuning  of the values of
of stochasticity coefficient (because
$D_{diff}=\frac{1}{4}(\frac{\varepsilon_0}{D})^2 T$)  and hence
allows a realization of different types of dynamics. 
We consider the following initial
conditions for (\ref{green}): a)
$\varepsilon=0.1$,~~~$S_z(0)=0.99,~~\varphi(0)=0$,~~~$T=0.01ps$, 
and $K_0=10^{-5}$. b)
$\varepsilon=0.1$,~~~$S_z(0)=0.99,~~\varphi(0)=0$,~~~$T=0.1ps$, and
$K_0=10^{-3}$.}\label{Fig:4}
\end{figure}
$$(\overline{s_{z}},\overline{\varphi})=\hat{T}(s_{z},\varphi),
\overline{s_{z}}=
s_{z}(t_{0}+T-0),$$
$$\overline{\varphi}=\varphi(t_{0}+T-0),\, s_{z}=
s_{z}(t_{0}-0),\, \varphi=\varphi(t_{0}-0).
$$
%\nonumber
%\end{eqnarray}
Note, that  $\hat{T}=\hat{T}_{R}\,
\hat{T}_{\delta}$  consists of two parts, one describing  the
free rotations $\hat{T}_{R}$ and  the other, $\hat{T}_{\delta}$,
 the action of  the applied pulses, i.e.
%\be\label{Rot}
$$\hat{T}_{R}(s_{z};\varphi)=(s_{z};\varphi+\omega(s_{z})T).$$
For $\hat{T}_{\delta}$
%integrating eq.(\ref{eq:motion}) on the small interval of time
%around the applied pulse $(t_{0}-0,t_{0}+0):$
%\begin{eqnarray}\label{eq:int}
we find upon integrating EOM
%%&&
%$s_{z}(t_{0}+0)-s_{z}(t_{0}-0)=\int\limits_{t_{0}-0}^{t_{0}+0}\dot{s}_{z}dt
%=-\int\limits_{t_{0}-0}^{t_{0}+0}\varepsilon\frac{\partial
%V_0(s_{z},\varphi)}{\partial
%\varphi}T\sum\limits_{k=-\infty}^{+\infty}\delta (t-k
%T)=-\varepsilon T\frac{\partial V_0(s_{z},\varphi)}{\partial
%\varphi}
%\varphi(t_0+0)-\varphi(t_0-0)=\int\limits_{t_{0}-0}^{t_{0}+0}\dot{\varphi}dt=\varepsilon
%T\frac{\partial V_0(\varphi),t}{\partial \varphi}.
%$
%%\end{eqnarray}
%As a result, we have \be \label{Te}
$$\hat{T}_\delta
(s_z,\varphi)=\bigg(s_z-\varepsilon\frac{\partial
V_0(s_{z},\varphi)}{\partial \varphi},\varphi+\varepsilon
T\frac{\partial V_0(s_{z},\varphi)}{\partial \varphi}\bigg).$$
%\ee
%%Combining (\ref{Rot}), (\ref{Te}) for total operator $\hat{T}$ we
%deduce:
%\begin{eqnarray}\label{st}
%&&(\bar{s}_z,\bar{\varphi})=\hat{T}(s_z,\varphi)=\hat{T}_R\hat{T}_\delta
%(s_z,\varphi)=\hat{T}_R\bigg(s_{z}-\varepsilon T\frac{\partial
%V_0(s_{z},\varphi)}{\partial \varphi},\varphi+\varepsilon
%T\frac{\partial V_0(s_{z},\varphi)}{\partial
%s_z}\bigg)=\nonumber\\&&=\bigg(s_z-\varepsilon T\frac{\partial
%V_0(s_{z},\varphi)}{\partial \varphi};\varphi +\omega(\bar{s}_z)T+
%\varepsilon T \frac{\partial V_0(s_{z},\varphi)}{\partial
%s_z}\bigg).
%\end{eqnarray}
Thus the recurrence relation applies
%\begin{eqnarray}\label{red}
%&&\
%$\bar{s_z}=s_z-\varepsilon T\frac{\partial
%V_0(s_{z},\varphi)}{\partial
%\varphi},\, \bar{\varphi}=\varphi +\omega(\bar{s}_z)T+
%\varepsilon T \frac{\partial V_0(s_{z},\varphi)}{\partial s_z}.$
%%\end{eqnarray}
%Or in the explicit form
\begin{eqnarray}\label{green}
&&\bar{s_z}=s_z-\varepsilon
T\sqrt{1-s_z^2}\sin\varphi,\\&&\bar{\varphi}_n=\varphi_n-s_z
T+\varepsilon T^2\sqrt{1-s_z^2}\sin\varphi+\frac{\varepsilon
Ts_z}{\sqrt{1-s_z^2}}\cos\varphi .\nonumber
\end{eqnarray}
Depending on the chosen parameters, these relations (\ref{green}) may be
be stable or unstable as signified by the corresponding  Lyapunov exponents.
Here, we inspect the Jacobean matrix $$
M=\frac{\partial (\overline{s_{z}},\overline{\varphi})}{\partial
(s_{z},\varphi)}=\left(\begin{array}{c}\frac{\partial
\overline{s_{z}}}{\partial s_{z}}~~~\frac{\partial
\overline{s_{z}}}{\partial \varphi}\\\frac{\partial
\overline{\varphi}}{\partial s_{z}}~~~\frac{\partial
\overline{\varphi}}{\partial \varphi}\end{array}\right)$$
and find for the eigenvalues
$$\lambda_{1,2}=1+\frac{1}{2}K\pm
\sqrt{(1+\frac{1}{2}K)^2-1},$$ $$ K=\varepsilon
T^2\frac{\partial^2V_0}{\partial\varphi^2}=\varepsilon
T^2\sqrt{1-s_z^2}\cos\varphi=K_0\sqrt{1-s_z^2}\cos\varphi,$$ where
$K_0=maxK=\varepsilon T^2.$
Chaos is expected if $\lambda_1>1$, i.e. for  $K>0$, meaning even
 weak pulses $K_0=\varepsilon T^2>0$ may lead to a diffusion.
  %This
%means that pulses, weaker than some critical values
%$K_0<K_0^{\prime}$ has no influence on the system  (see Fig.4). We
%see if chaos is realized then dynamics of
 I.e., $S_n^z(t)=S^z(t=nT)$
and the magnetization can be deflected diffusively and irreversibly
if
$K_0$ exceeds a critical value $K_0^{\prime}$,
as demonstrated by the numerical calculations in Fig.\ref{Fig:3}.
Essential for this phenomena is the
 existence of two time scales, the slow variables
$S_n^z(t)$ and the fast random phase $\varphi(t)$.
Using  the random phase
approximation for the fast phases \cite{Toklikishvili} one infers
the  Fokker-Planck equation
%\be\label{presto}
$$ \frac{\partial
f(s_z,t)}{\partial t}=\frac{1}{2}\frac{\partial}{\partial
s_z}D(s_z)\frac{\partial f(s_z,t)}{\partial s_z},
$$
%\ee
where $$D(s_z)=\frac{\pi\varepsilon^2}{\Omega}\sum\limits_{m=-\infty}^{+\infty}m^2|V_m(s_z)|^2~\Omega=\frac{2\pi}{T};$$
and $V_m(s_z)$ are the Fourier  coefficients as deduced from the expansion
$$V_0(s_z,\varphi)=\sum\limits_{n=-\infty}^{+\infty}V_m(s_z)\exp(in\varphi).$$
Explicitly we have
$$V_0(s_z,\varphi)=-\sqrt{1-s_z^2}\cos\varphi=\frac{\sqrt{1-s_z^2}}{2}(e^{i\varphi}+e^{-i\varphi}).$$
Therefore, $$V_1=V_{-1}=-\frac{\sqrt{1-s_z^2}}{2},$$
and the diffusion coefficient is
% \be \label{free}
$$D(s_z)=\frac{\pi\varepsilon^2}{\Omega}(|V_{-1}|^2+|V_1|^2)=\frac{\pi\varepsilon^2}{2\Omega}(1-s_z^2)=
\frac{\varepsilon^2T}{4}(1-s_z^2).$$
%\ee
 Consequently, we write
 %\be \label
%{ball}
$$ \frac{\partial f(s_z,t)}{\partial
t}=D_{diff}\frac{\partial}{\partial s_z}(1-s_z^2)\frac{\partial
f(s_z,t)}{\partial s_z},$$
%\ee
where $$D_{diff}=\frac{\varepsilon^2
T}{4}$$  is the coefficient of diffusion which is completely
defined by the pulse parameters \be \label{fast}\frac{d}{dt}\langle
S_z\rangle=-2D\langle S_z\rangle,\, \langle
S_z\rangle=\int\limits_{-1}^{+1}S_zf(S_z,t)dS_z.\ee  Thus, the mean value
of the spin projection behaves as  $$ \langle S_z(t)\rangle=\langle
S_z(0)\rangle e^{-2D_{diff}t}.$$
Since the diffusion coefficient is given by
$$D_{diff}=\frac{\varepsilon^2
T}{4}=\frac{1}{4}\left(\frac{\varepsilon_0}{D}\right)^2 T$$ ($D$ is the magnetic anisotropy constant)
$D_{diff}$ can be tuned by changing appropriately the external field parameters, e.g.
 by varying the amplitude of the pulses $\varepsilon_0$ and/or 
the interval  $T$ between them. Depending on these parameters   different types of
the dynamics is realized.  To test for this analytical prediction
we performed full numerical calculations that are in good accord with 
 the analytical results (see Fig.\ref{Fig:4}). This statement is based on
 the fact that
for
$\varepsilon=0.1,~T=0.1,~D_{diff}=\varepsilon^2T=0.001$ the
analytically estimated decay rate $1/D_{diff}$ coincides with the
numerically deduced one (cf. Fig.\ref{Fig:4}).

\section{Summary}
The aim of this work is to  point out 
the possibility of 
a stochastic energy pumping and magnetization deflection
in a single molecular magnet subjected to a static and a time-variable,
polychromatic magnetic fields.
The key point is that the parameters of the applied
static and pulsed magnetic fields can be tuned such that the system
is driven nearby a separatrix where the magnetization dynamics turns  diffusive
allowing thus for a magnetization switching even in the absence of damping (that conventionally originates
from coupling to other degrees of freedom).

\textbf{Acknowledgment:} The  project is  supported by
the Georgian National Foundation (grants: GNSF/STO 7/4-197,
GNSF/STO 7/4-179) and by the Deutsche
Forschungsgemeinschaft (DFG) through SFB 762 and through  SPP 1285.

\end{document}